\documentclass[letterpaper]{article} 
\usepackage{aaai2027}  
\nocopyright  
\usepackage[hyphens]{url}  
\usepackage{graphicx} 
\usepackage{natbib}  
\usepackage{caption} 
\usepackage{amssymb}
\usepackage{amsmath}
\usepackage{tabularx}
\usepackage{graphicx}
\usepackage{multirow}
\usepackage{orcidlink}
\usepackage{algorithm}
\usepackage{algorithmic}

\usepackage{newfloat}
\usepackage{listings}
\DeclareCaptionStyle{ruled}{labelfont=normalfont,labelsep=colon,strut=off} 
\floatstyle{ruled}
\newfloat{listing}{tb}{lst}{}
\floatname{listing}{Listing}

\usepackage{booktabs}

\title{
Exploring Joint-Embedding Predictive Architectures for Self-Supervised Representation Learning of Powder X-ray Diffraction Patterns
}
\author{
    Ming-Chiang Chang\textsuperscript{\rm 1}\corresponding\thanks{\orcidlink{0000-0003-1880-6654}0000-0003-1880-6654},
    Brian DeCost\textsuperscript{\rm 1}\corresponding\thanks{\orcidlink{0000-0002-3459-5888}0000-0002-3459-5888}
}
\affiliations{
    \textsuperscript{\rm 1}National Institute of Standards and Technology\\
    100 Bureau Dr\\
    Gaithersburg, MD 20899 USA\\
    mingchiang.chang@nist.gov\\
    brian.decost@nist.gov
}

\begin{document}

\maketitle

\begin{abstract}
Powder X-ray diffraction (XRD) is the primary high-throughput method for resolving crystal structures, a cornerstone of materials science. As high-throughput synthesis and self-driving laboratories advance, automated XRD pattern analysis has become vital for accelerating materials discovery. However, traditional matrix factorization methods fail when lattice variations cause non-linear peak shifts. At the same time, supervised learning approaches face a combinatorial explosion when handling multi-phase mixtures. To overcome these challenges, we adopt a self-supervised Joint Embedding Predictive Architecture (JEPA) to learn representation embeddings from 1D diffraction patterns. By optimizing a masked self-prediction loss directly within the latent space, the framework captures high-level crystallographic semantics. Downstream evaluations show that lightweight attentive probes trained on frozen JEPA representations achieve space group classification and lattice parameter estimation accuracy close to much larger supervised vision transformers.
Furthermore, latent probing reveals that the learned representation space naturally organizes physical transformations, isolating smooth trajectories for peak broadening.
Interestingly, decoding along latent paths between single-phase crystal representations yields realistic multi-phase mixture patterns, demonstrating an emergent continuous structure learned without explicit multi-phase supervision.

\end{abstract}

\section{Introduction}
New and better materials have been the driving force for the advancement and prosperity of societies.
The emergence of self-driving laboratories represents the next major catalyst in accelerating materials discovery and development~\cite{alab, cameo, sarah}.
Still, key technical hurdles remain before these facilities can truly achieve autonomy and the speedup they wish to accomplish.
Chief among them is X-ray diffraction (XRD) analysis.
While XRD is the primary method for resolving crystalline structures, yielding rich insights into grain size, microstrain, and secondary phases, interpreting these patterns programmatically remains remarkably challenging~\cite{xrd}.
As illustrated in Figure \ref{schematic}, an XRD pattern consists of discrete peaks whose positions reflect interplanar lattice spacings and whose intensities correspond to atomic species and arrangements within the unit cell.
However, automated analysis is complicated by the nonlinear shift in peak positions caused by linear lattice variations, as well as the challenge of deconvoluting multi-phase mixture spectra.
The former precludes the direct application of standard matrix factorization techniques like nonnegative matrix factorization (NMF)~\cite{nmf}, while the latter creates a combinatorial explosion of phase combinations that severely complicates supervised learning.
As most previous work approaches the XRD analysis problem with supervised learning on specific prediction tasks~\cite{oviedo2019fast, lee2020deep, szymanski2021probabilistic}, self-supervised representation learning of XRD patterns is underexplored.

Recently, joint embedding predictive architecture (JEPA) was proposed for self-supervised representation learning, achieving this by predicting masked regions with provided context within a learned latent space~\cite{jepa, jepa_concept}.
Supervision based on latent space predictions forces JEPA to capture high-level semantic representations rather than low-level pixel details, reducing computational overhead and enabling higher-level reasoning.
Followup works extend this concept to deal with videos and develop a world model for robotic controls and planning~\cite{v-jepa2, leworldmodel}.
Other than computer vision and robotics, JEPA has also been widely adapted to serve other purposes, such as audio~\cite{a-jepa}, genomics~\cite{dnajepa}.

Applying self-supervised representation learning, and JEPA in particular, to XRD analysis is uniquely compelling.
Traditional generative or reconstructive self-supervised models, such as Masked Autoencoders (MAE), force the network to reconstruct exact 1D intensity profiles, causing the model to overfit to background noise or minor peak displacements~\cite{mae,subramanian2025xxact}.
Conversely, contrastive learning forces discrete clustering via negative pairs, creating sharp latent boundaries rather than a smooth, continuous representation space.~\cite{jepa_concept, wang2020understanding}.
JEPA circumvents these limitations: by performing predictions purely in the latent space, it enables the model to learn invariant crystallographic features across continuous lattice shifts and complex peak overlaps without needing expensive pixel-level reconstructions or artificial augmentations.

In this work, we demonstrate the application of a self-supervised JEPA model to build an information-rich representation space for XRD patterns.
We show that the self-supervised pre-training trajectory is highly stable, with downstream performance scaling consistently as representation learning progresses. 
Leveraging the learned JEPA representations, lightweight probe networks achieve high accuracy in downstream crystallographic tasks, such as space group classification and lattice parameter estimation, comparable to much larger ResNet~\cite{resnet} and vision transformer (ViT)~\cite{vit} models, while drastically reducing computational costs and data requirements relative to fully supervised models.
JEPA also demonstrates superior performance compared to MAE, which we attribute to its latent-space prediction strategy that avoids expending model capacity on reconstructing low-level pixel noise.
Furthermore, vector arithmetic within the latent space reveals semantically meaningful directions that map directly to continuous peak broadening variations.
Finally, we observe that interpolating between latent representations of different crystal structures produces valid mixture XRD patterns upon decoding, demonstrating an emergent capability not explicitly included in the training data.
Ultimately, this work establishes JEPA as a powerful foundational paradigm for XRD analysis, bridging the gap toward fully autonomous materials characterization in next-generation self-driving laboratories.

\section{Method}
\begin{figure}[t]
    \centering
    \includegraphics[width=1.0\columnwidth]{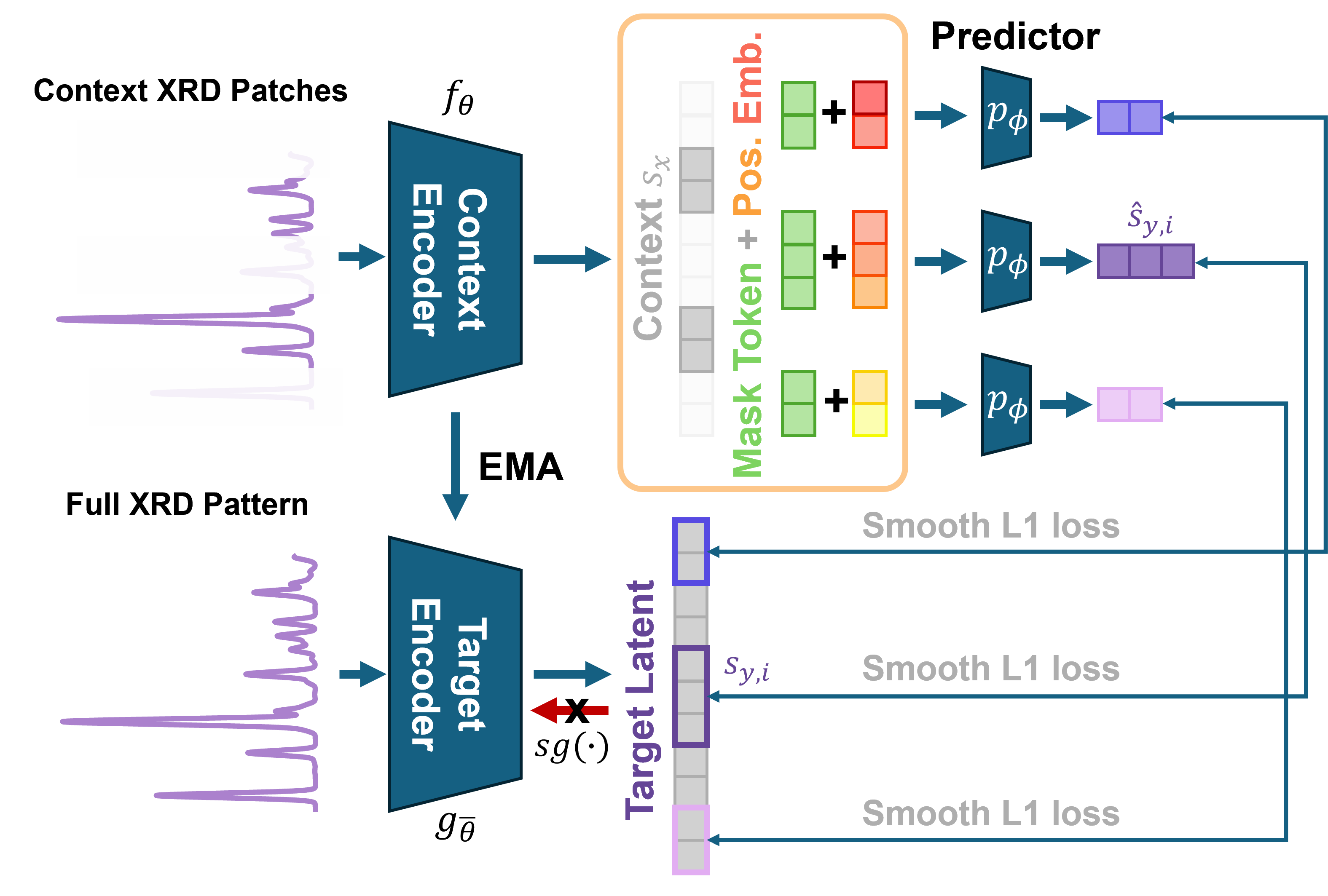} 
    \caption{Schematic of the JEPA model architecture trained on XRD patterns, illustrating the context encoder, predictor, and target encoder pipelines.}
    \label{schematic}
\end{figure}

\paragraph{JEPA model training}
The model architecture schematics and the training workflow are shown in Fig. \ref{schematic}.
The JEPA learns self-supervised representations by predicting target latent features from context input, avoiding pixel- or token-level reconstruction and guiding the model to learn high-level semantic representations.

First, given a crystal structure $c$, stochastic physical simulation and augmentations $\mathcal{T}$ (detailed later in this section) are first applied to obtain $\tilde{x} = \mathcal{T}(c)$.
This augmented pattern $\tilde{x}$ is partitioned into a sequence of $N$ non-overlapping patches.

To create the prediction task, we sample a context region $M_x \subset \{1, \dots, N\}$ using a retention ratio $r \sim \mathcal{U}(0.85, 1.0)$ and designate $K$ target patch regions $M_{y, i} \subset \{1, \dots, N\}$ ($K=2$ in this work).
Context patches in $M_x$ overlapping with target regions ${M_y= \bigcup_i M_{y,i}}$ are omitted prior to encoding.
For the remaining context patches, sinusoidal positional embeddings are applied using their absolute indices in the original grid.
The context patches are then processed by the context encoder $f_\theta$ to produce context representations $s_x \in \mathbb{R}^{\vert M_x \setminus M_y \vert \times d}$.
Concurrently, the target encoder $g_{\bar{\theta}}$ processes the full sequence $E$, and target latents $s_{y, i} = g_{\bar{\theta}}(E)[M_{y, i}] \in \mathbb{R}^{\vert M_{y, i} \vert \times d}$ are extracted by indexing the target patch positions $M_y$.

Before feeding the context representation into the predictor network $p_\phi$, target patch positions are substituted with a learned predictor mask token $p_{\text{mask}} \in \mathbb{R}^d$, and positional embeddings $P$ are re-added to restore spatial coordinates.
The predictor takes the concatenated context tokens and target-positioned mask tokens to estimate target representations:
\begin{equation}
    \hat{s}_{y, i} = p_\phi\Big(\big[s_x + P[M_x],\; p_{\text{mask}} + P[M_{y, i}]\big]\Big)[\vert{}M_x\vert{}:] \in \mathbb{R}^{\vert{}M_{y, i}\vert{} \times d}
\end{equation}
where $P[M_x]$ and $P[M_{y, i}]$ injects the spatial coordinates of the context and target patches, respectively.
Finally, the context encoder weights $\theta$ and predictor weights $\phi$ are jointly updated by minimizing the Smooth $L_1$ loss in latent space across target positions:
\begin{equation}
    \mathcal{L}_{\text{JEPA}}(\theta, \phi) = \frac{1}{|M_y|} \sum_i^K \text{Smooth}_{L1}\left( \hat{s}_{y, i} - \text{sg}\left(s_{y, i}\right) \right),
\end{equation}
where $\text{sg}(\cdot)$ represents the stop-gradient operator. For an element-wise prediction residual vector $e = \hat{s}_{y, i} - \text{sg}(s_{y, i}) \in \mathbb{R}^d$, the $\text{Smooth}_{L1}$ function is defined as:
\begin{equation}
    \text{Smooth}_{L1}(e) = \frac{1}{d} \sum_{j=1}^{d} \begin{cases} 
        0.5 \, e_j^2 & \text{if } |e_j| < 1, \\[6pt]
        |e_j| - 0.5 & \text{otherwise.}
    \end{cases}
\end{equation}
To prevent representation collapse without needing negative samples, target encoder parameters $\bar{\theta}$ are updated asynchronously as an exponential moving average (EMA) of $\theta$.
During training, the momentum coefficient was initialized to $m = 0.9999$ and linearly increased to $m = 1.0$ by the end of training.

We employ a symmetric pair of identical ViT architectures for the context and target encoders, coupled with a lightweight ViT as the predictor.
In actual application, both target encoder and the predictor are dropped and we only use the context encoder for generating XRD representation.
In this work, we use the embedding dimension of 128 with 4 head within each multi-head attention module for the JEPA encoders and predictors.
Two JEPA models were trained with encoder depths ($d$) of 4 and 8 blocks, paired with predictor depths of 2 and 4 blocks, respectively.
The benchmark models use an input dimension of 384, whereas the decoder-equipped model uses an input dimension of 768 to enhance visual representation, and both models use a patch size of 16.
For baseline comparisons, we evaluate standard ViT variants (ViT-Tiny, ViT-Small, and ViT-Base \cite{vit, deit}) alongside ResNet-50 \cite{resnet}, adhering to their standard architecture configurations.
The self-supervised models are trained for 1000 epochs and the attentive probing is trained for 300 epochs, same as the baseline models.

\paragraph{Dataset and Augmentation}
We use the crystal structures from the MP-20 dataset used in \citet{cdvae} and use the same train, validation, test split.
Powder XRD patterns are simulated over a scattering vector magnitude ($q$) ranging from 8 to $60\text{ nm}^{-1}$.
To ensure that the learned representations remain robust to experimental variations and structural distortions, we apply three stochastic, physics-grounded augmentations during pre-training:

\begin{enumerate}
    \item \textbf{Symmetry-Preserving Strain:} To model physical lattice deformations while adhering to fundamental crystallographic constraints, we construct diagonal strain tensors $\mathbf{\varepsilon} = \text{diag}(\varepsilon_a, \varepsilon_b, \varepsilon_c)$ without shear components. Normal strain components are sampled uniformly as $\varepsilon_i \sim \mathcal{U}(-0.1, 0.1)$. The tensor components are constrained according to the crystal system of each material (e.g., $\varepsilon_a = \varepsilon_b = \varepsilon_c$ for cubic systems, and $\varepsilon_a = \varepsilon_b \neq \varepsilon_c$ for tetragonal and hexagonal systems), inducing physically realistic, non-linear shifts in peak location via Bragg's law.

    \item \textbf{Peak Broadening:} To approximate finite crystallite size and instrumental broadening effects, diffraction peaks are convolved with a Gaussian smoothing kernel. The standard deviation $\sigma_{\text{width}}$ of the Gaussian kernel is sampled heuristically from a uniform distribution $\mathcal{U}(0.05, 0.3)$~nm$^{-1}$, dynamically varying peak widths across generated patterns.

    \item \textbf{Background Noise:} Real-world diffraction patterns inherently contain sample fluorescence, diffuse scattering, and counting statistics noise.
    In this exploratory work, we inject background artifacts by adding Gaussian noise $\eta \sim \mathcal{N}(0, \sigma^2)$ directly to the pattern intensities:
    \begin{equation}
        I_{\text{aug}}(q) = I(q) + \eta.
    \end{equation}
    In our training, $\sigma~\sim~\mathcal{U}(0, 0.03)$ after the intensity of the XRD pattern being normalized to have max of one.
\end{enumerate}

\paragraph{Downstream Adaptation and Attentive Probing}
We evaluate the quality of the JEPA representations by performing attentive probing, following \citet{v-jepa2}.
The pre-trained encoder remains entirely frozen, while a lightweight cross-attention module aggregates the patch embeddings.
Specifically, a set of learnable query tokens (analogous to a classification [\texttt{CLS}] token~\cite{vit}) attends to the patch embeddings, pooling them into a compact global representation.
Finally, a lightweight multilayer perceptron (MLP) maps this representation to the target prediction.

In this study, we apply the learned JEPA representations to predict the space group and lattice parameters directly from XRD patterns.
Space group classification determines the underlying crystal symmetry, whereas lattice parameter regression quantifies unit cell dimensions.
Together, they form the core structural identity required to model and design functional materials.
Lattice parameters are well known to be sensitive to the choice of unit cell, and the most commonly used standard unit cell convention (Niggli reduction~\cite{niggli1929}) can exhibit sharp discontinuities with respect to small symmetry-breaking distortions~\cite{andrews2019space}.
To mitigate this, our model predicts the $S^6$ parameterization of the Selling-reduced unit cell, and we use the symmetrized $S^6$ distance formulated by \citet{andrews2019space} to properly measure the prediction power provided by the learned representation.

To evaluate the latent representations learned by JEPA and explore downstream applications, we train a pixel reconstruction decoder to reconstruct input XRD patterns directly from the JEPA embeddings.
This decoder mirrors the architecture of the JEPA encoder, adding a shallow MLP after the final attention block to project the embedding back to XRD pattern patches.

\section{Results}
\subsection{Representation Quality Evolution}

\begin{figure}[t]
    \centering
    \includegraphics[width=0.9\columnwidth]{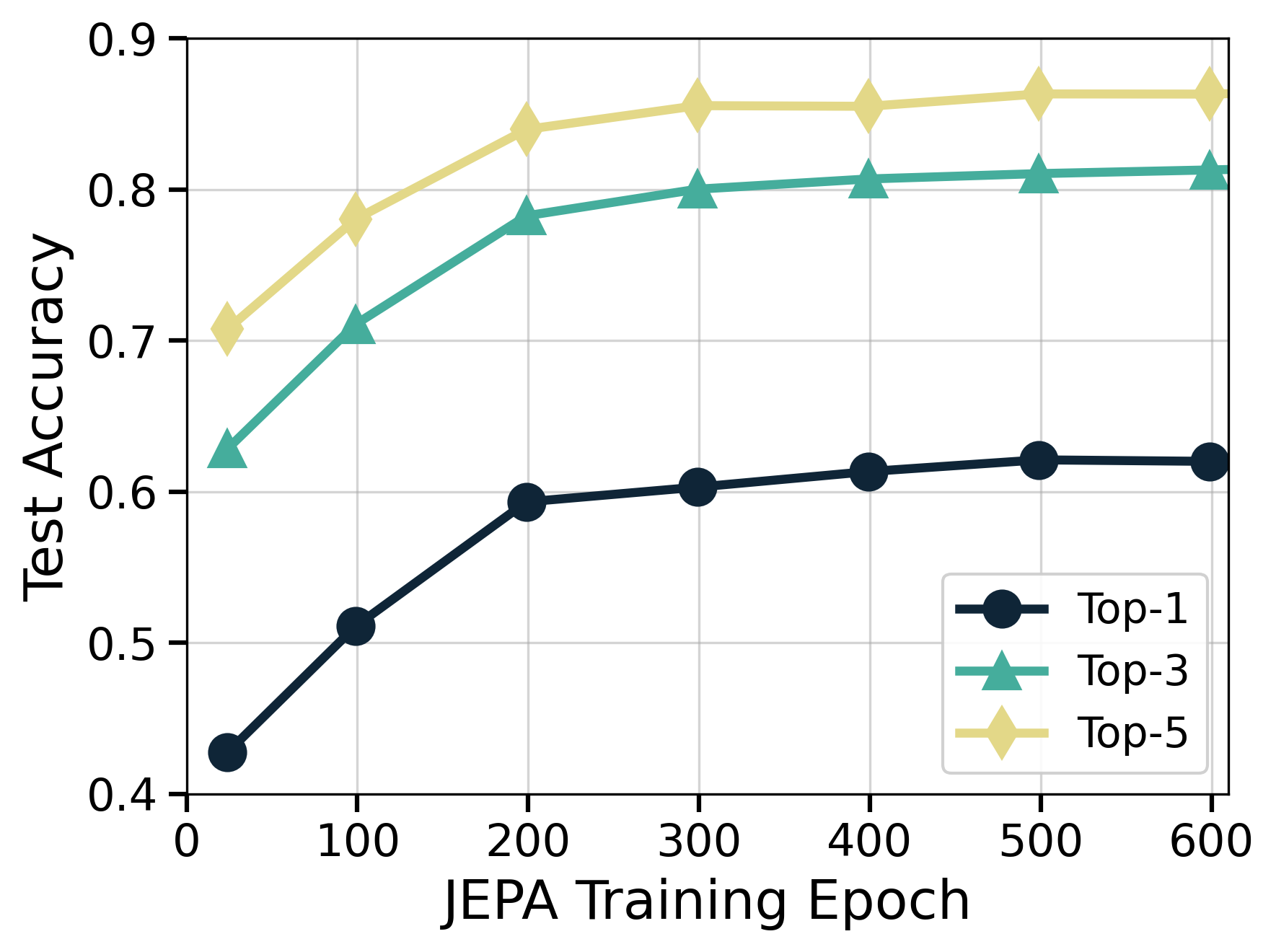} 
    \caption{Downstream space group prediction performance trajectories using attentive probing across JEPA training epoch checkpoints.
    Accuracy on space group prediction scales consistently with representation learning progress, demonstrating stable feature convergence.}
    \label{representation_evolution}
\end{figure}
To understand how self-supervised representation learning gains knowledge in the domain that can directly translate to downstream task performance, we tracked attentive probing accuracy on space group prediction throughout JEPA training epochs (Figure~\ref{representation_evolution}).
Probing performance was evaluated on the MP-20 test set for space group classification.
As shown in Figure~\ref{representation_evolution}, accuracy improves steadily with longer JEPA training across all metrics (Top-1, Top-3, and Top-5), indicating continuous refinement of the learned XRD representations.
Rather than hitting an early bottleneck, the model continuously unlocks richer crystallographic representations as training progresses, driving Top-1 accuracy up to nearly 62~\% and Top-5 accuracy beyond 85~\%.
This steady, monotonic improvement up to 600 epochs demonstrates training stability and confirms that JEPA representation learning captures the subtle, complex physical signatures required for high-accuracy space group identification.
Paired with downstream accuracy comparable much larger architectures (as demonstrated later in this section), this trajectory indicates that the JEPA framework effectively encodes rich, parameter-efficient representations from raw XRD patterns.

\begin{table*}[t]
\centering
\small
\begin{tabular}{llc|cc|c}
  \toprule
  & \textbf{Model / Protocol} & \textbf{Trainable Params} & \textbf{Top-1 Acc (\%)} & \textbf{Top-5 Acc (\%)} & \textbf{Lattice Error (\AA)} \\
\midrule
  & & & \multicolumn{2}{c|}{\textbf{Space Group Classification }} & $S^6$ \textbf{Distance} \\
  \midrule
  \multirow{6}{*}{\rotatebox{90}{\textit{Supervised}}}
  & 1D-ResNet-50   & 16$\times 10^6$  & $63.27 \pm 2.20$ & $86.44 \pm 0.86$ & $\textbf{0.345} \pm \textbf{0.009}$  \\
  & ViT-Tiny       & 5$\times 10^6$   & $61.82 \pm 0.31$ & $85.67 \pm 0.30$ & $0.501 \pm 0.012$ \\
  & ViT-Small      & 22$\times 10^6$  & $\underline{66.92 \pm 0.46}$ & $87.96 \pm 0.25$ & $0.448 \pm 0.015$ \\
  & ViT-Base       & 86$\times 10^6$  & $\textbf{68.39} \pm \textbf{0.15}$  & $\textbf{88.54} \pm \textbf{0.10}$ & $\underline{0.365 \pm 0.023}$ \\
  & ViT-Special (d=4) & $0.8\times10^6$ & $46.60 \pm 0.71$ & $74.66 \pm 0.82$ & $0.510 \pm 0.001$\\
  & ViT-Special (d=8) & $1.6\times10^6$ & $52.11 \pm 0.77$ & $79.72 \pm 0.26$ & $0.496 \pm 0.001$\\
  \midrule
  \multirow{4}{*}{\rotatebox{90}{\textit{SSL}}}
  & MAE + Attentive Probe (d=4) & (0.8+0.2)$\times 10^6$  & $61.13 \pm 0.30$ & $85.82 \pm 0.14$ & $0.421 \pm 0.003$ \\
  & MAE + Attentive Probe (d=8) & (1.6+0.2)$\times 10^6$  & $62.65 \pm 0.61$ & $86.63 \pm 0.26$ & $0.413 \pm 0.002$ \\
  & \textbf{JEPA + Attentive Probe (d=4)} & (0.8+0.2)$\times 10^6$  & $63.84 \pm 0.50$ & $87.30 \pm 0.21$ & $0.410 \pm 0.011$ \\
  & \textbf{JEPA + Attentive Probe (d=8)} & (1.6+0.2)$\times 10^6$  & $65.22 \pm 0.17$ & $\underline{88.12 \pm 0.21}$ & $0.387 \pm 0.002$ \\
\end{tabular}
\caption{Comparative evaluation of downstream performance between JEPA and baseline architectures across crystallographic classification tasks.
Each task compartment highlights performance and parameter efficiency against fully fine-tuned baseline architectures.
Bold and underline denote the best and second-best performing models, respectively.
The lattice error is computed as the minimal Euclidean distance between two unit cells in $S_6$ space across all 168 Selling-equivalent permutations, accounting for symmetry boundary reflections~\cite{andrews2019space}.
The standard deviation is from five repeated runs with different seed.
For each metric, bold text highlights the top performer and underlined text denotes the second-best performer.}
\label{tab:comparison}
\end{table*}

\subsection{Crystallography Prediction Task Adaptation}

\begin{figure}[t]
    \centering
    \includegraphics[width=\columnwidth]{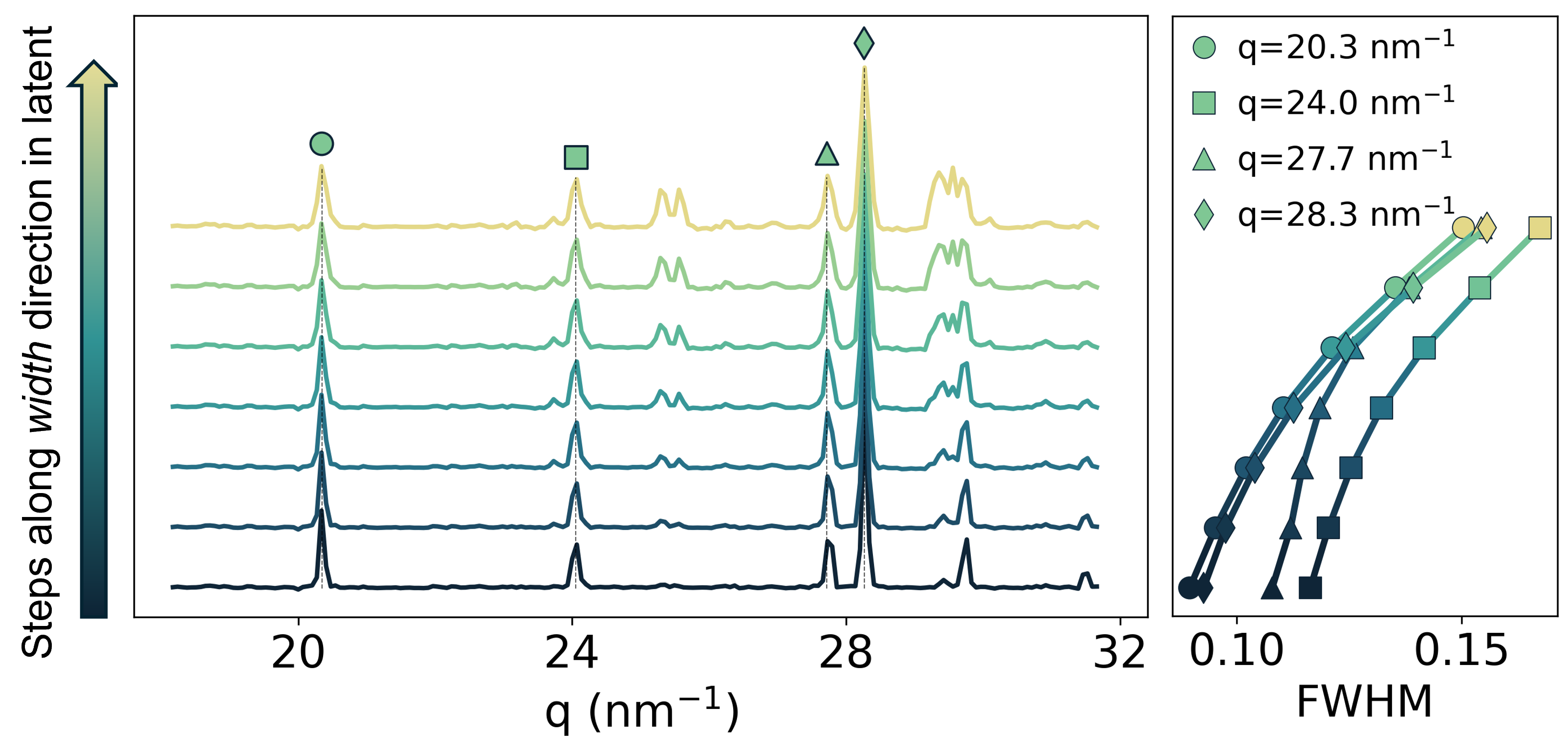} 
    \caption{Left: A series of reconstructed XRD patterns along the discovered width direction in the latent space. 
    The bottom curve shows the reconstructed XRD pattern obtained by encoding a known crystal structure.
    Each pattern above it moves in the \textit{width} direction extracted by PCA in the latent before decoder reconstruction.
    Right: fitted Gaussian peak widths of four principal peaks along the latent trajectory.
    }
    \label{width}
\end{figure}
We evaluate the performance of our framework on the 230-class space group classification benchmark against supervised convolutional model and transformer baselines as summarized in Table \ref{tab:comparison}.
Fully supervised architectures exhibit standard capacity scaling, progressing from 61.82~\% Top-1 accuracy for ViT-Tiny (5 million parameters) to 68.39~\% for ViT-Base (86 million parameters).
Our JEPA model, evaluated under an attentive probing protocol with a frozen encoder, achieves 65.22~\% Top-1 and 88.12~\% Top-5 accuracy with depth of 8.
Notably, JEPA outperforms the fully supervised ViT-Tiny baseline and ResNet, despite operating with an ultra-compact 1.6 million parameter backbone and only 0.2 million trainable probe parameters.

While heavily parameterized supervised models like 1D-ResNet-50 (16 million parameters) and ViT-Base (86 million parameters) achieve higher Top-1 accuracy, they rely on full end-to-end backpropagation across tens of millions of task-specific weights.
In contrast, our JEPA-based model adds merely 0.2 million parameters during probing, demonstrating that the self-supervised pre-training objective successfully yields physically meaningful representations of crystal symmetries without label supervision.
Performance gains when scaling encoder depth from $d=4$ to $d=8$ highlight the architecture's capacity for parameter scaling.
Notably, our JEPA variants employ a substantially narrower embedding dimension than standard ViTs; demonstrating stable convergence at higher depths and wider architecture underscores their potential for efficient, larger-scale pretraining.

\begin{figure*}[t]
    \centering
    \includegraphics[width=1.4\columnwidth]{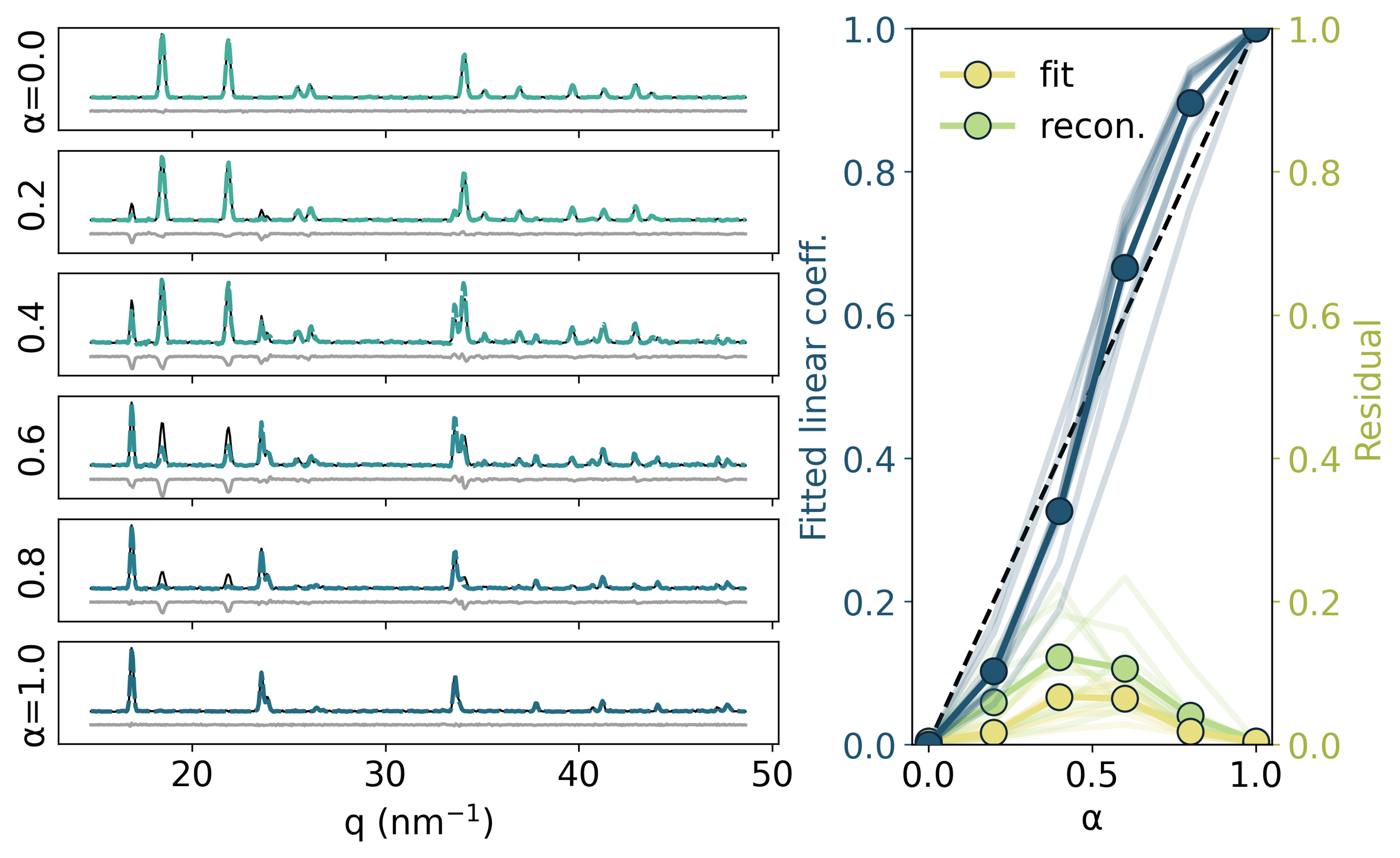} 
    \caption{Left: Interpolated XRD patterns for two rutile structures.
    Colored dashed lines represent patterns decoded from the interpolated latent vectors, solid black lines show the direct linear mixtures of the experimental XRD patterns, and gray lines indicate the residual differences ($\alpha$ denotes the mixture coefficient).
    Right: Statistics of the linear analysis from decoding interpolated latent vectors.
     The blue lines represent linear mixture coefficients fitted using endpoint XRD patterns against decoded mixtures.
     Light yellow lines show the residuals from the linear fit (capturing decoding artifacts), while light green lines show the overall residual between reconstructed and mixture XRD patterns (accounting for both nonlinearity and decoding artifacts).
     Transparent lines denote individual trajectories across the 10 structure pairs, while solid lines indicate their mean.
     The close alignment with the ideal dashed line demonstrates strong linear mixing behavior, with both residuals peaking as expected at a 50:50 mixture ($\alpha = 0.5$).
     Both residuals are defined as $\sum_i (y_i - \hat{y}_i)^2 / \sum_i y_i^2$.
     For the reconstruction residual, $y$ is the ground-truth mixture pattern and $\hat{y}$ is the decoded pattern. For the fitting residual, $y$ is the decoded pattern and $\hat{y}$ is the linear fit derived from the two single-phase end-point XRD patterns.
    }
    \label{fig:interpolate}
\end{figure*}

Lattice parameter prediction presents a unique challenge, as models must deduce structural constants by capturing both long-range relationships and precise peak positions.
Surprisingly, in our evaluation, the 1D-ResNet-50 yields the lowest error ($0.345$~\AA), demonstrating that deep convolution-based architectures remain highly effective at learning global context and long-range dependencies.
On the other hand, Vision Transformers scale effectively with model capacity, progressively lowering the error from $0.501~\AA$ ($\text{ViT-Tiny}$, 5 million parameters) down to $0.365~\AA$ ($\text{ViT-Base}$, 86 million parameters).
Notably, the JEPA model paired with a lightweight attentive probe achieves a prediction loss of $0.410~\AA$ and $0.387~\AA$ for $d=4$ and $d=8$, respectively, significantly surpassing the performance of the supervised ViT-Small while requiring much smaller parameter footprint.

To compare representations learned from latent-space versus pixel-level reconstruction, we also trained MAEs using the same ViT backbones.
Across both space group prediction and lattice parameter regression tasks, the JEPA models consistently outperformed their MAE counterparts.
We attribute this performance gain to JEPA's architectural design: by predicting in the latent space, JEPA avoids expending model capacity on reconstructing low-level pixel noise, thereby encouraging the encoder to preserve high-level semantic representations and yielding superior downstream prediction performance at equivalent model scale.

Finally, we evaluate a baseline trained directly in a supervised manner on the same ViT architecture used for our self-supervised models (denoted as ViT-Special in Table \ref{tab:comparison}).
These supervised models yield substantially lower performance compared to their self-supervised counterparts, demonstrating that self-supervised representation learning is far more parameter-efficient and represents a promising direction for future work.

Since JEPA demonstrates strong representational capacity relative to its compact parameter footprint and ViT demonstrates strong improvement in scaling, it is reasonable to assume that scaling would greatly enhance JEPA's performance and may surpass ViTs on downstream tasks with a larger, deeper model.
However, scaling the JEPA architecture poses non-trivial optimization challenges.
Because self-supervised training relies on EMA updates to decouple the context and target encoders, maintaining stability and preventing representation collapse requires delicate hyperparameter tuning.
More recent formulations, such as LeJEPA~\cite{lejepa}, offer promising strategies to circumvent EMA updates, potentially facilitating more straightforward scaling.
We defer model scaling to future work and turn instead to analyzing the latent space learned through JEPA training.

\subsection{Semantic Directions in JEPA Latent}
In this section, we analyze the physical interpretability and structural organization of the JEPA latent space.
To isolate the latent direction corresponding to peak broadening, we first generated a synthetic series of XRD patterns derived from a crystal structure in the MP-20 dataset, systematically increasing the peak widths across the sequence.
After embedding these patterns using the pre-trained JEPA encoder, we performed principal component analysis (PCA) on the resulting latent representations and defined the primary principal component as the characteristic "width direction". 
To evaluate whether this vector generalizes, we selected another crystal structure, embedded its baseline XRD pattern, and incrementally stepped along the identified width direction in the latent space.
Decoding these perturbed embeddings back into XRD patterns yields a semantically coherent trajectory of transformed spectra, as illustrated in the left panel Figure~\ref{width}. 
Moving along this latent vector monotonically increases peak broadening while preserving overall reflection positions and relative intensities, albeit introducing minor artifacts such as spurious peaks near $q = 25\text{ nm}^{-1}$.
The right panel of Figure~\ref{width} provides the full width at half maximum (FWHM) of the fitted Gaussian distribution of four principal peaks as the latent vector traverses along the peak-width direction, showing a consistent, monotonic increase in peak broadening.
This continuous transformation demonstrates that the JEPA encoder autonomously discovers physically grounded, disentangled representations without explicit supervision.
Beyond enhancing interpretability, this latent controllability paves the way for downstream models operating directly within the latent space, leveraging its structured semantic organization for advanced XRD analysis.
For comparison, we also examined whether MAE encoders develop similar directional semantics in their latent spaces, but the evidence is weaker compared to JEPA models (Figure \ref{fig:mae_vs_jepa} in the Appendix).
This hints toward an advantage of latent-space reconstruction over pixel reconstruction for semantic extraction, though we defer a thorough quantification to future work.

Note that utilizing this technique for lattice strain did not demonstrate similar behavior.
This suggests that lattice variations are not linearly encoded in the latent space, but instead form a non-linear local manifold.

\subsection{Emergent Behavior for Mixtures}
One major hurdle of XRD analysis is the combinatorial problem when more than one phase exists.
Previous research primarily relies on either brute-force combinatorial training, expensive tree search, or subtract-and-detect searching, both of which present distinct challenges~\cite{lee2020deep, chang2025probabilistic, szymanski2021probabilistic, Fei2026}.
Direct supervised training on synthetic mixtures scales exponentially with the number of candidate phases, rendering exhaustive dataset generation computationally intractable.
Alternatively, guided tree search becomes computationally prohibitive when exploring large design spaces.
Finally, sequential subtract-and-detect routines suffer from severe error propagation, where minor fitting inaccuracies in primary phases distort residual signals and trigger downstream misidentifications.

To bypass these combinatorial bottlenecks, analyzing patterns directly within a continuous latent space offers a promising alternative.
However, for a latent space to reliably model physical phase mixtures, a representation capable of comprehending linear superpositions is desirable.
Here, we demonstrate that the JEPA model, despite being trained strictly on single-phase XRD patterns, can learn to reconstruct mixtures of XRD patterns from its latent in a systematic and physically consistent way.
As shown in Figure~\ref{fig:interpolate} (Left), interpolating between the latent vectors of two distinct rutile structures yields decoded XRD patterns (colored dashed lines) that closely mirror direct linear combinations of the endpoint patterns (solid black lines).
To evaluate mixture behavior across the latent path, we randomly sampled ten pairs of crystal structures, performed latent interpolation between their representations, and fitted the decoded patterns against linear combinations of the endpoint spectra (Figure~\ref{fig:interpolate}, Right).
The blue lines represent the normalized linear coefficients of the endpoint mixtures that best fit the decoded patterns across all ten structure pairs, with the solid blue line indicating the mean trajectory.
his coefficient increases monotonically following the ideal diagonal line, exhibiting a slight sigmoidal profile that reflects a nonlinear phase influence: the secondary phase contribution grows gradually near the source anchor, accelerates through the midpoint, and saturates near the target anchor.
The overall residual between the reconstructed and target mixture patterns (light green curve in Fig.~\ref{fig:interpolate}) remains generally low, peaking slightly above $0.1$ near equal phase proportions ($\alpha = 0.5$).
However, this total residual conflates geometric non-linearities in the latent path with genuine decoding artifacts.
To isolate the decoding artifacts from non-linear path effects, we evaluate the residual from the linear fits alone, shown as the light yellow curve in Figure~\ref{fig:interpolate}.
Crucially, the linear fitting residual is significantly lower than the reconstruction residual, especially around $\alpha=0.5$.
This provides evidence that decoding along the latent path introduces minimal artifacts and yields patterns closely resemble mixture patterns, supporting the hypothesis that the learned latent space naturally organizes multi-phase representations along continuous transition paths without requiring multi-phase training data.
Similar behavior was reported by \citet{gomez6889867variational} in variational autoencoders, and we observed comparable behaviors in MAEs (Figure \ref{fig:interpolate_mae} in the Appendix), highlighting an intriguing phenomenon that warrants further investigation and development.

\section{Conclusion}
In this work, we explored the potential of JEPA pretraining for XRD data analysis.
By framing representation learning strictly within the latent space, this approach bypasses limitations of traditional pixel- or intensity-reconstruction methods, offering robust handling of experiment artifacts.
The learned embeddings provide an information-rich, physically grounded foundation that enables downstream models to achieve favorable predictive accuracy with extremely compact models, making it suitable for ultrafast XRD pattern screening.
By using principal component analysis, a latent direction related to peak broadening can be identified, showing that self-supervised learning can encode physical semantics without explicit supervision.
We also demonstrated an emergent behavior, where a mixture pattern can be reconstructed by a trained decoder by interpolating between two crystal representations, a property that, to our knowledge, has not been clearly demonstrated in self-supervised framework before.
Together, these findings highlight how JEPA's self-supervised representations can bridge the gap between deep learning and physical interpretability, paving the way for scalable, real-time XRD analysis.

\pagebreak
\appendix
\renewcommand{\thefigure}{A.\arabic{figure}}
\setcounter{figure}{0} 
\section{Appendix}
\label{sec:appendix_visualizations}

\begin{figure}[H] 
  \centering
  
  \includegraphics[width=\columnwidth]{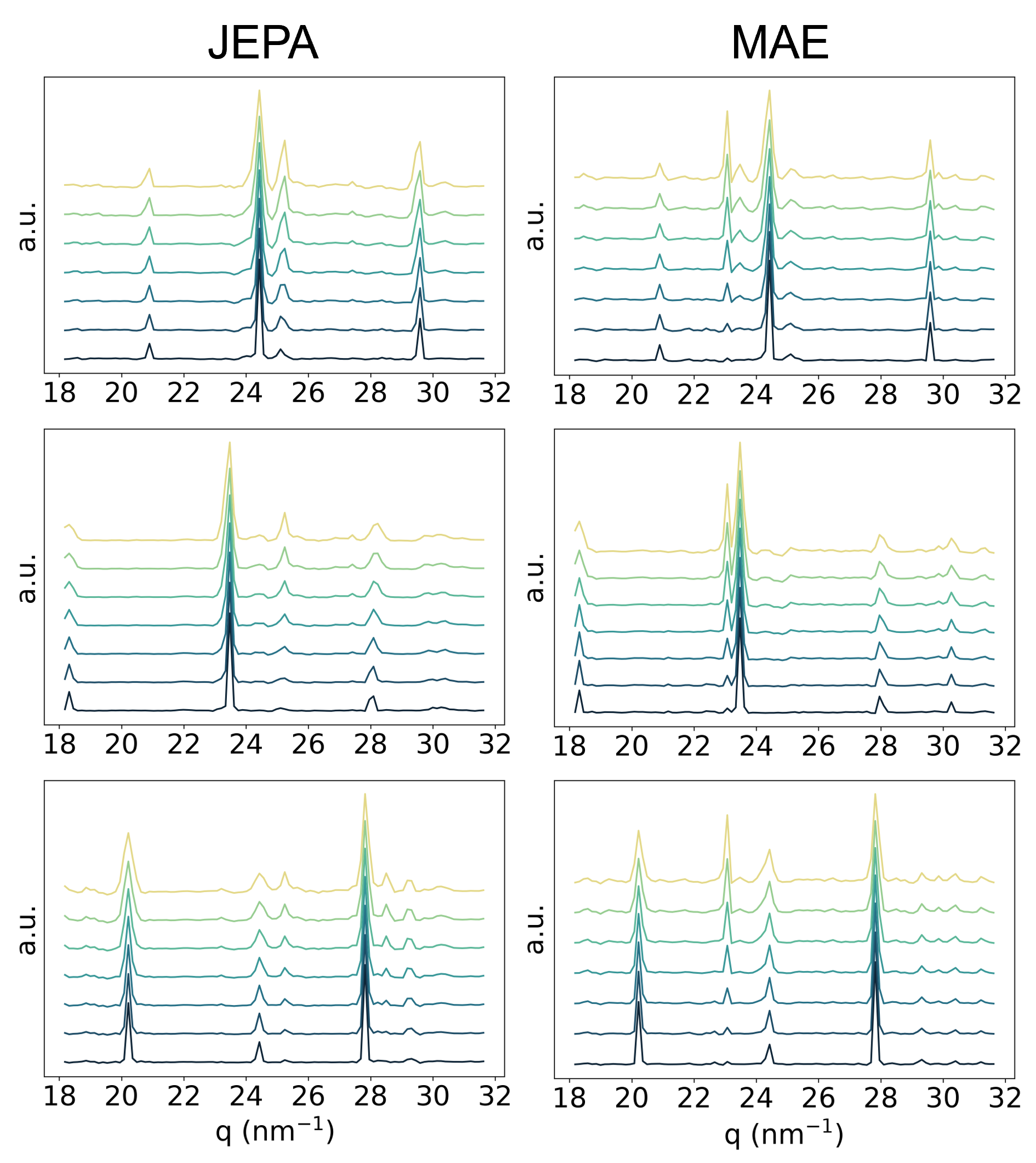}
  \caption{Qualitative comparison of XRD pattern reconstructions along the extracted latent width trajectory starting from the same seed crystal.
    Each panel displays a sequence of XRD patterns reconstructed by moving along the discovered width direction in latent space, starting from the encoded representation of a known crystal structure (bottom curve).
    Overall, the JEPA model produces more consistent and physically plausible peak broadening along the trajectory, whereas the MAE reconstructions exhibit inconsistent trends and produce more artifacts.
    A quantitative evaluation of these latent trajectories is deferred to future work.}
  \label{fig:mae_vs_jepa}
  
  \vspace{0.2cm} 
  
  \includegraphics[width=\columnwidth]{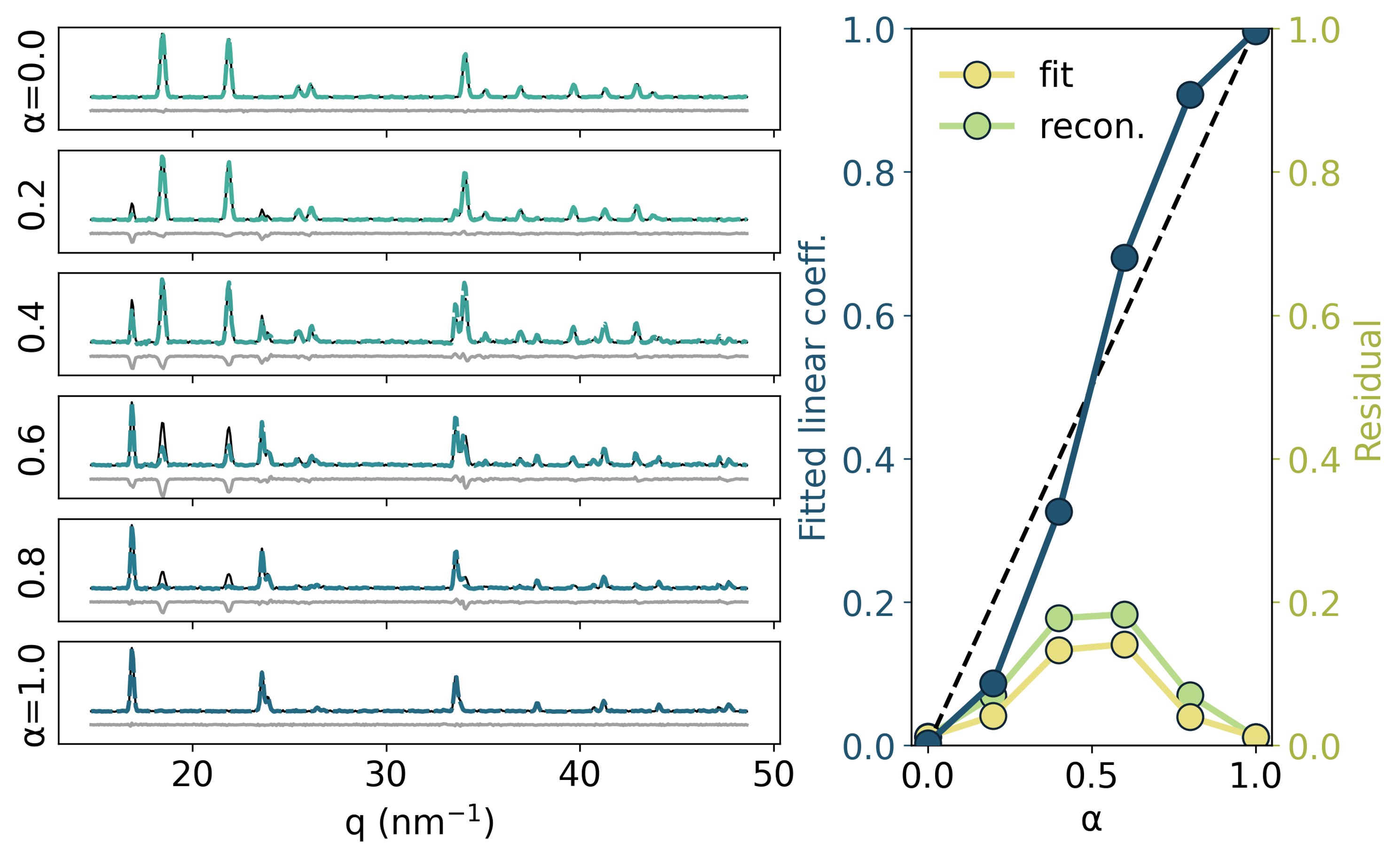}
  \caption{Latent space interpolation using the Masked Autoencoder (MAE) decoder.
    XRD patterns were reconstructed by linearly interpolating between latent representations in the MAE model, evaluated using the same decoding protocol as JEPA.
    }
  \label{fig:interpolate_mae}
\end{figure}

\section*{Acknowledgments}
Certain equipment, instruments, software, or materials are identified in this paper in order to specify the adequately. 
Such identification is not intended to imply recommendation or endorsement of any product or service by NIST, nor is it intended to imply that the materials or equipment identified are necessarily the best available for the purpose.
The computational results in this work were made possible by the Redwing high performance computer cluster on NIST's Boulder campus and supported by NIST's Research Data and Computing Office. 
This work was funded in part by the NIST Innovation in Measurement Science (IMS) program.


\begin{thebibliography}{27}
\providecommand{\natexlab}[1]{#1}

\bibitem[{Andrews, Bernstein, and Sauter(2019)}]{andrews2019space}
Andrews, L.~C.; Bernstein, H.~J.; and Sauter, N.~K. 2019.
\newblock A space for lattice representation and clustering.
\newblock \emph{Foundations of Crystallography}, 75(3): 593--599.

\bibitem[{Assran et~al.(2025)Assran, Bardes, Fan, Garrido, Howes, Muckley,
  Rizvi, Roberts, Sinha, Zholus et~al.}]{v-jepa2}
Assran, M.; Bardes, A.; Fan, D.; Garrido, Q.; Howes, R.; Muckley, M.; Rizvi,
  A.; Roberts, C.; Sinha, K.; Zholus, A.; et~al. 2025.
\newblock V-jepa 2: Self-supervised video models enable understanding,
  prediction and planning.
\newblock \emph{arXiv preprint arXiv:2506.09985}.

\bibitem[{Assran et~al.(2023)Assran, Duval, Misra, Bojanowski, Vincent, Rabbat,
  LeCun, and Ballas}]{jepa}
Assran, M.; Duval, Q.; Misra, I.; Bojanowski, P.; Vincent, P.; Rabbat, M.;
  LeCun, Y.; and Ballas, N. 2023.
\newblock Self-supervised learning from images with a joint-embedding
  predictive architecture.
\newblock In \emph{2023 IEEE/CVF Conference on Computer Vision and Pattern
  Recognition (CVPR)}, 15619--15629. IEEE.

\bibitem[{Balestriero and LeCun(2025)}]{lejepa}
Balestriero, R.; and LeCun, Y. 2025.
\newblock Lejepa: Provable and scalable self-supervised learning without the
  heuristics.
\newblock \emph{arXiv preprint arXiv:2511.08544}.

\bibitem[{Chang et~al.(2025)Chang, Ament, Amsler, Sutherland, Zhou, Gregoire,
  Gomes, van Dover, and Thompson}]{chang2025probabilistic}
Chang, M.-C.; Ament, S.; Amsler, M.; Sutherland, D.~R.; Zhou, L.; Gregoire,
  J.~M.; Gomes, C.~P.; van Dover, R.~B.; and Thompson, M.~O. 2025.
\newblock Probabilistic phase labeling and lattice refinement for autonomous
  materials research.
\newblock \emph{npj Computational Materials}, 11(1): 148.

\bibitem[{Chang et~al.(2026)Chang, Amsler, Sutherland, Ament, Gann, Zhou,
  Smieska, Woll, Gregoire, Gomes, van Dover, and Thompson}]{sarah}
Chang, M.-C.; Amsler, M.; Sutherland, D.~R.; Ament, S.; Gann, K.~R.; Zhou, L.;
  Smieska, L.~M.; Woll, A.~R.; Gregoire, J.~M.; Gomes, C.~P.; van Dover, R.~B.;
  and Thompson, M.~O. 2026.
\newblock Autonomous Materials Exploration Integrates Automated Phase
  Identification and AI Agents Enhanced by Human Guidance.
\newblock \emph{PRX Intelligence}, 1: 013019.

\bibitem[{Cullity and Smoluchowski(1957)}]{xrd}
Cullity, B.~D.; and Smoluchowski, R. 1957.
\newblock Elements of X-ray Diffraction.
\newblock \emph{Physics Today}, 10(3): 50--50.

\bibitem[{Dosovitskiy et~al.(2020)Dosovitskiy, Beyer, Kolesnikov, Weissenborn,
  Zhai, Unterthiner, Dehghani, Minderer, Heigold, Gelly et~al.}]{vit}
Dosovitskiy, A.; Beyer, L.; Kolesnikov, A.; Weissenborn, D.; Zhai, X.;
  Unterthiner, T.; Dehghani, M.; Minderer, M.; Heigold, G.; Gelly, S.; et~al.
  2020.
\newblock An image is worth 16x16 words: Transformers for image recognition at
  scale.
\newblock \emph{arXiv preprint arXiv:2010.11929}.

\bibitem[{Fei et~al.(2026)Fei, McDermott, Rom, Wang, and Ceder}]{Fei2026}
Fei, Y.; McDermott, M.~J.; Rom, C.~L.; Wang, S.; and Ceder, G. 2026.
\newblock Dara: Automated Multiple-Hypothesis Phase Identification and
  Refinement from Powder X-ray Diffraction.
\newblock \emph{Chemistry of Materials}, 38(3): 1364–1376.

\bibitem[{Fei, Fan, and Huang(2023)}]{a-jepa}
Fei, Z.; Fan, M.; and Huang, J. 2023.
\newblock A-jepa: Joint-embedding predictive architecture can listen.
\newblock \emph{arXiv preprint arXiv:2311.15830}.

\bibitem[{G{\'o}mez~Peralta, Bokhimi, and
  Quintana-Owen(2026)}]{gomez6889867variational}
G{\'o}mez~Peralta, J.~I.; Bokhimi, X.; and Quintana-Owen, P. 2026.
\newblock Variational autoencoders for generating multiphase powder diffraction
  patterns and their quantification.
\newblock \emph{Available at SSRN 6889867}.

\bibitem[{He et~al.(2022)He, Chen, Xie, Li, Doll{\'a}r, and Girshick}]{mae}
He, K.; Chen, X.; Xie, S.; Li, Y.; Doll{\'a}r, P.; and Girshick, R. 2022.
\newblock Masked autoencoders are scalable vision learners.
\newblock In \emph{2022 IEEE/CVF conference on computer vision and pattern
  recognition (CVPR)}, 15979--15988. IEEE.

\bibitem[{He et~al.(2016)He, Zhang, Ren, and Sun}]{resnet}
He, K.; Zhang, X.; Ren, S.; and Sun, J. 2016.
\newblock Deep residual learning for image recognition.
\newblock In \emph{Proceedings of the IEEE conference on computer vision and
  pattern recognition}, 770--778.

\bibitem[{Kusne et~al.(2020)Kusne, Yu, Wu, Zhang, Hattrick-Simpers, DeCost,
  Sarker, Oses, Toher, Curtarolo et~al.}]{cameo}
Kusne, A.~G.; Yu, H.; Wu, C.; Zhang, H.; Hattrick-Simpers, J.; DeCost, B.;
  Sarker, S.; Oses, C.; Toher, C.; Curtarolo, S.; et~al. 2020.
\newblock On-the-fly closed-loop materials discovery via Bayesian active
  learning.
\newblock \emph{Nature communications}, 11(1): 5966.

\bibitem[{Larey et~al.(2026)Larey, Dahan, Bleiweiss, Kellerman, Leib, Nayshool,
  Ofer, Zinger, Dominissini, Rechavi et~al.}]{dnajepa}
Larey, A.; Dahan, E.; Bleiweiss, A.; Kellerman, R.; Leib, G.; Nayshool, O.;
  Ofer, D.; Zinger, T.; Dominissini, D.; Rechavi, G.; et~al. 2026.
\newblock Jepa-dna: Grounding genomic foundation models through joint-embedding
  predictive architectures.
\newblock \emph{arXiv preprint arXiv:2602.17162}.

\bibitem[{LeCun et~al.(2022)}]{jepa_concept}
LeCun, Y.; et~al. 2022.
\newblock A path towards autonomous machine intelligence version 0.9. 2,
  2022-06-27.
\newblock \emph{Open Review}, 62(1): 1--62.

\bibitem[{Lee and Seung(1999)}]{nmf}
Lee, D.~D.; and Seung, H.~S. 1999.
\newblock Learning the parts of objects by non-negative matrix factorization.
\newblock \emph{nature}, 401(6755): 788--791.

\bibitem[{Lee et~al.(2020)Lee, Park, Lee, Singh, and Sohn}]{lee2020deep}
Lee, J.-W.; Park, W.~B.; Lee, J.~H.; Singh, S.~P.; and Sohn, K.-S. 2020.
\newblock A deep-learning technique for phase identification in multiphase
  inorganic compounds using synthetic XRD powder patterns.
\newblock \emph{Nature communications}, 11(1): 86.

\bibitem[{Maes et~al.(2026)Maes, Lidec, Scieur, LeCun, and
  Balestriero}]{leworldmodel}
Maes, L.; Lidec, Q.~L.; Scieur, D.; LeCun, Y.; and Balestriero, R. 2026.
\newblock Leworldmodel: Stable end-to-end joint-embedding predictive
  architecture from pixels.
\newblock \emph{arXiv preprint arXiv:2603.19312}.

\bibitem[{Niggli(1929)}]{niggli1929}
Niggli, P. 1929.
\newblock Krystallographische und strukturtheoretische Grundbegriffe (Handbuch
  der Experimentalphysik, Bd VII, 1).
\newblock \emph{Geologiska F\"{o}reningen i Stockholm F\"{o}rhandlingar},
  51(1): 121–122.

\bibitem[{Oviedo et~al.(2019)Oviedo, Ren, Sun, Settens, Liu, Hartono, Ramasamy,
  DeCost, Tian, Romano et~al.}]{oviedo2019fast}
Oviedo, F.; Ren, Z.; Sun, S.; Settens, C.; Liu, Z.; Hartono, N. T.~P.;
  Ramasamy, S.; DeCost, B.~L.; Tian, S.~I.; Romano, G.; et~al. 2019.
\newblock Fast and interpretable classification of small X-ray diffraction
  datasets using data augmentation and deep neural networks.
\newblock \emph{npj Computational Materials}, 5(1): 60.

\bibitem[{Subramanian et~al.(2025)Subramanian, Hung, Schweigert, Suram, and
  Ye}]{subramanian2025xxact}
Subramanian, J.; Hung, L.; Schweigert, D.; Suram, S.; and Ye, W. 2025.
\newblock XxaCT-NN: Structure Agnostic Multimodal Learning for Materials
  Science.
\newblock \emph{arXiv preprint arXiv:2507.01054}.

\bibitem[{Szymanski et~al.(2021)Szymanski, Bartel, Zeng, Tu, and
  Ceder}]{szymanski2021probabilistic}
Szymanski, N.~J.; Bartel, C.~J.; Zeng, Y.; Tu, Q.; and Ceder, G. 2021.
\newblock Probabilistic deep learning approach to automate the interpretation
  of multi-phase diffraction spectra.
\newblock \emph{Chemistry of Materials}, 33(11): 4204--4215.

\bibitem[{Szymanski et~al.(2023)Szymanski, Rendy, Fei, Kumar, He, Milsted,
  McDermott, Gallant, Cubuk, Merchant et~al.}]{alab}
Szymanski, N.~J.; Rendy, B.; Fei, Y.; Kumar, R.~E.; He, T.; Milsted, D.;
  McDermott, M.~J.; Gallant, M.; Cubuk, E.~D.; Merchant, A.; et~al. 2023.
\newblock An autonomous laboratory for the accelerated synthesis of inorganic
  materials.
\newblock \emph{Nature}, 624(7990): 86--91.

\bibitem[{Touvron et~al.(2021)Touvron, Cord, Douze, Massa, Sablayrolles, and
  J{\'e}gou}]{deit}
Touvron, H.; Cord, M.; Douze, M.; Massa, F.; Sablayrolles, A.; and J{\'e}gou,
  H. 2021.
\newblock Training data-efficient image transformers \& distillation through
  attention.
\newblock In \emph{International conference on machine learning}, 10347--10357.
  PMLR.

\bibitem[{Wang and Isola(2020)}]{wang2020understanding}
Wang, T.; and Isola, P. 2020.
\newblock Understanding contrastive representation learning through alignment
  and uniformity on the hypersphere.
\newblock In \emph{International conference on machine learning}, 9929--9939.
  PMLR.

\bibitem[{Xie et~al.(2022)Xie, Fu, Ganea, Barzilay, and Jaakkola}]{cdvae}
Xie, T.; Fu, X.; Ganea, O.-E.; Barzilay, R.; and Jaakkola, T.~S. 2022.
\newblock Crystal Diffusion Variational Autoencoder for Periodic Material
  Generation.
\newblock In \emph{International Conference on Learning Representations}.

\end{thebibliography}
\end{document}